\documentclass[slac_one]{revtex4}
\usepackage{graphicx}
\usepackage{fancyhdr}
\newcommand{\ee}  {\ensuremath{ e^+e^- }}
\newcommand{\sqs} {\ensuremath{ \sqrt{s} }}
\newcommand{\cc}  {\ensuremath{ c\bar{c} }}

\newcommand{\gisr}{\ensuremath{ \gamma_{\mathrm{ISR}}}}
\newcommand{\gtot}{\ensuremath{ \Gamma_{\mathrm{tot}}}}

\newcommand{\gev}  {\ensuremath{\, {\text{GeV}}}}
\newcommand{\mev}  {\ensuremath{\, {\text{MeV}}}}
\newcommand{\gevc} {\ensuremath{\, {\text{GeV}}/c^2}}
\newcommand{\mevc} {\ensuremath{\, {\text{MeV}}/c^2}}
\newcommand{\st}   {\ensuremath{\, {\text{stat.}}}}
\newcommand{\sys}  {\ensuremath{\, {\text{sys.}}}}

\newcommand{\jp}  {\mbox{\ensuremath{  J/\psi}}}
\newcommand{\pp}  {\mbox{\ensuremath{\psi(2S)}}}

\newcommand{\pipi}    {\ensuremath{\pi^+\pi^-}}
\newcommand{\ppjpsi}  {\ensuremath{\pi^+\pi^- J/\psi  }}
\newcommand{\pppsip}  {\ensuremath{\pi^+\pi^- \psi(2S)}}

\newcommand{\dpdstm}   {\ensuremath{ D^+ D^{*-}     }}
\newcommand{\dstpdstm} {\ensuremath{ D^{*+} D^{*-}  }}
\newcommand{\dndmpp}   {\ensuremath{ D^0 D^- \pi^+ }}

\newcommand{\dd}       {\ensuremath{ D   \overline D      }}
\newcommand{\ddst}     {\ensuremath{ D   \overline D {}^* }}
\newcommand{\dstdst}   {\ensuremath{ D^* \overline D {}^* }}

\newcommand{\dnd}     {\ensuremath{ D^0    \overline D {}^{0}}}
\newcommand{\dndst}   {\ensuremath{ D^0    \overline D {}^{*0}}}

\newcommand{\eedd}     {\ensuremath{ e^+e^- \to D \overline D }}

\newcommand{\lala} {\ensuremath{ \Lambda_c^+  \Lambda_c^- }}
\newcommand{\mll}  {\ensuremath{ M_{\Lambda_c^+  \Lambda_c^-}}}
\newcommand{\eell} {\ensuremath{ e^+e^- \to \Lambda_c^+  \Lambda_c^- }}

\begin{document}

\title{Exotic $c \overline c$ spectroscopy}
\author{G.Pakhlova}
\affiliation{ITEP, Moscow, Russia}
\begin{abstract}
We review the recent experimental results on the exotic charmoniumlike
states.  Among them we discuss the $X(3872)$, $Y(3940)$, $Z^\pm(4430)$
and $Z^\pm_{1,2}$ states found in $B$-meson decays, the $X(3940)$ and
$X(4160)$ states produced in double charmonium production and the
$Y(4260)$, $Y(4325)$, $Y(4660)$ and $X(4630)$ states produced with
initial-state radiation in \ee\ annihilation.
\end{abstract}
\maketitle
\thispagestyle{fancy}

\section{INTRODUCTION}

The first charmonium state $\jp(1S)$, the bound system consisting of
the charmed quark $c$ and anti-quark $\overline c$, was discovered in
1974~\cite{jps}. Nine more charmonium states, the $\eta_c(1S)$,
$\chi_{c0}(1P)$, $\chi_{c1}(1P)$, $\chi_{c2}(1P)$, \pp, $\psi(3770)$,
$\psi(4040)$, $\psi(4160)$ and $\psi(4415)$ were observed shortly
afterwards. Some of them, the so called $\psi$ states with quantum
numbers $J^{PC}=1^{--}$, were found in \ee\ annihilation. Four
observed $\psi$ resonances have masses above open charm
threshold~\footnote{mass of two charm mesons $\sim 3.73\gevc$.}. Other
states were observed in radiative decays of the \jp\ or \pp. During
the next two decades no other charmonium states were found. Meanwhile,
the decay properties of the known states were carefully studied. By
the beginning of the XXI century the masses and widths of \cc\ states
as well as their hadron, lepton and two-photon decays were described
by theory with good accuracy.

A new charmonium era started in 2002. During the past six years
numerous charmoniumlike states were discovered. Among them, only the
$h_c(1P)$~\cite{cleo:hc}, $\eta_c(2S)$~\cite{belle:etac2s} and
$Z(3930)\equiv \chi_{c2}(2P)$~\cite{belle:z3930} have been identified
as candidates for conventional charmonium, while a number of other
states with masses above open charm threshold have serious problems
with a charmonium interpretation. Many states observed in $B$ decays
remain unidentified: the $X(3872)$~\cite{belle:x3872},
$Y(3940)$~\cite{belle:y3940} and $Z^\pm(4430)$,
$Z^{\pm}_{1,2}$~\cite{belle:z4430,belle:z12}. The nature of the so
called $Y(4260)$, $Y(4325)$, $Y(4660)$ and $X(4630)$
states~\cite{babar:y4260,belle:y4260,babar:y4360,belle:y4360,belle:x4630}
with quantum numbers $J^{PC}=1^{--}$ that are produced in
\ee\ annihilation also remains unclear. Two new states, the
$X(3940)$~\cite{belle:x3940} and $X(4160)$~\cite{belle:x4160}, were
discovered in double charmonium production and have also not yet been
assigned to any of the vacant charmonium levels.

There are a variety of theoretical interpretations of the new
states. The conservative models~\cite{thresh} suggest a
reconsideration the effect of the numerous open charm thresholds on
the parameters of the conventional \cc\ states predicted within the
potential models. However, most recent approaches admit the existence
of exotic states in the charmoniumlike spectrum. Among them are: the
models suggesting multiquark states that include either molecular
states (two loosely bound charm mesons
$(c\overline{q})(\bar{c}q)$)~\cite{molecular}, or tetraquarks (tightly
bound four-quark states,
$[cq][\overline{c}\overline{q}]$)~\cite{diquark}; charmonium hybrids
(\cc-gluon with excited gluonic degrees of freedom)~\cite{hybrid};
hadro-charmonium (compact charmonium states, $J/\psi$, $\psi(2S)$,
$\chi_c$, "coated" by excited light-hadron matter)~\cite{hadroch}. An
indication of the existence of exotic states would be the observation
of charmoniumlike states with quantum numbers forbidden for
conventional charmonium or an extremely narrow width. Discovery of
states with non-zero charge (e.g. $[cu][\overline{cd}]$), strangeness
(e.g.  $[cd][\overline{cs}]$) or both (e.g. $[cu][\overline{cs}]$)
would be also an obvious sign of the existence of multiquark states.

%%%%%%%%%%%%%%%%%%%%%%%%%%%%%%%%%%%%%%%%%%%%%%%%%%%%%%%%%%%%%%%%%%%%%

\section{$X(3872)$}

The narrow charmoniumlike state $X(3872)$ produced in the exclusive
decay $B^\pm \to K^\pm \ppjpsi$ was discovered by Belle in
2003~\cite{belle:x3872} with a statistical significance of
$10.3\sigma$. The mass of this state, which decays in \ppjpsi, was
measured to be $(3872.0\pm 0.6(\st) \pm 0.5(\sys))\mevc$ in a close
proximity to the $M_{D^0}+M_{D^{*0}}$ mass threshold. The width of the
$X(3872)$ was found to be surprisingly small: $\Gamma<2.3\mev$ at the
$90\%$ C.L.  The existence of the $X(3872)$ was confirmed by
CDF~\cite{CDF:x3872} and D0~\cite{D0:x3872}, who reported the
observation of a \ppjpsi\ resonance consistent with $X(3872)$ produced
inclusively in $p\overline p$ collisions, and by
BaBar~\cite{babar:x3872} who found $X(3872)$ in $B^\pm \to K^\pm
\ppjpsi$ decays. Today the world average mass and width values of the
$X(3872)$ are $M = (3872.2 \pm 0.8)\mevc$ and $\Gamma =
(3.0^{+2.1}_{-1.7})\mev$~\cite{pdg08}, respectively.

It was found by Belle~\cite{belle:x3872} and confirmed by
CDF~\cite{CDF:x3872_rho} that the \pipi\ invariant masses concentrate
near the upper kinematic boundary that corresponds to the $\rho^0$
meson mass. Charmonium decays to $\rho^0 \jp$ violate isospin and are
expected to be strongly suppressed. Observation of the decay
$X(3872)\to \jp \gamma$~\cite{babar:x3872_gam,belle:x3872_gam} and
indications of the decay $X(3872)\to \omega \jp$ fix $C_X=+1$ and
confirm that the decay $X(3872)\to \ppjpsi$ proceeds via $\rho^0
\jp$. Spin-parity analysis of the $X(3872)$ state in the final state
$\mu^+\mu^-\pipi$ performed by CDF~\cite{CDF:x3872_JPC} demonstrated
that only $C$-even assignments $J^{PC}=1^{++}$ and $2^{-+}$, decay via
$\jp\rho^0$ in the both cases, describe the data. Belle
measurements~\cite{belle:x3872_JPC} favor quantum numbers
$J^{PC}=1^{++}$. However, the $J>1$ assignments were not examined in
the Belle angular analysis.

Neither the $\chi_{c1}(2P)$ (corresponding to $J^{PC}=1^{++}$) nor the
$\eta_{c2}$ (corresponding to $J^{PC}=2^{-+}$) are expected to have
such a large branching fraction for the decay to the isospin violating
$\rho^0 \jp$ mode. For example, the $\eta_{c2}$ is expected to decay
dominantly into light hadrons, while the $\chi_{c1}(2P)$ should have a
${\cal B}(X(3872)\to \jp \gamma)$ that is at least two orders of
magnitude larger than that measured for the $X(3872)$. Moreover, the
mass of the $X(3872)$ is $\sim 100\mevc$ smaller than the expected
$\chi_{c1}(2P)$ mass. The most popular option for the $X(3872)$
interpretation is an S-wave \dndst\ molecular
state~\cite{x3872:molecule}. This proposal is motivated by the
proximity of the $X(3872)$ to the \dndst\ threshold: $M_X\sim
M_{D^0}+M_{D^{*0}}=(3871.81 \pm 0.25)\mevc$~\cite{pdg08}. In the
molecular model, the $X(3872)$ is naturally a $J^{PC}=1^{++}$ state;
admixtures of decays to $\rho^0 \jp$ and $\omega \jp$ are expected; a
small ${\cal B}(X(3872)\to \jp \gamma)$ is also predicted. Other
options for the $X(3872)$ are tetraquark states~\cite{x3872:tetra},
hybrids~\cite{x3872:hybrid} or a threshold effects~\cite{x3872:cusp}.

\begin{figure*}[t]
\centering
\begin{tabular}{cc}
\includegraphics[width=0.47\textwidth]{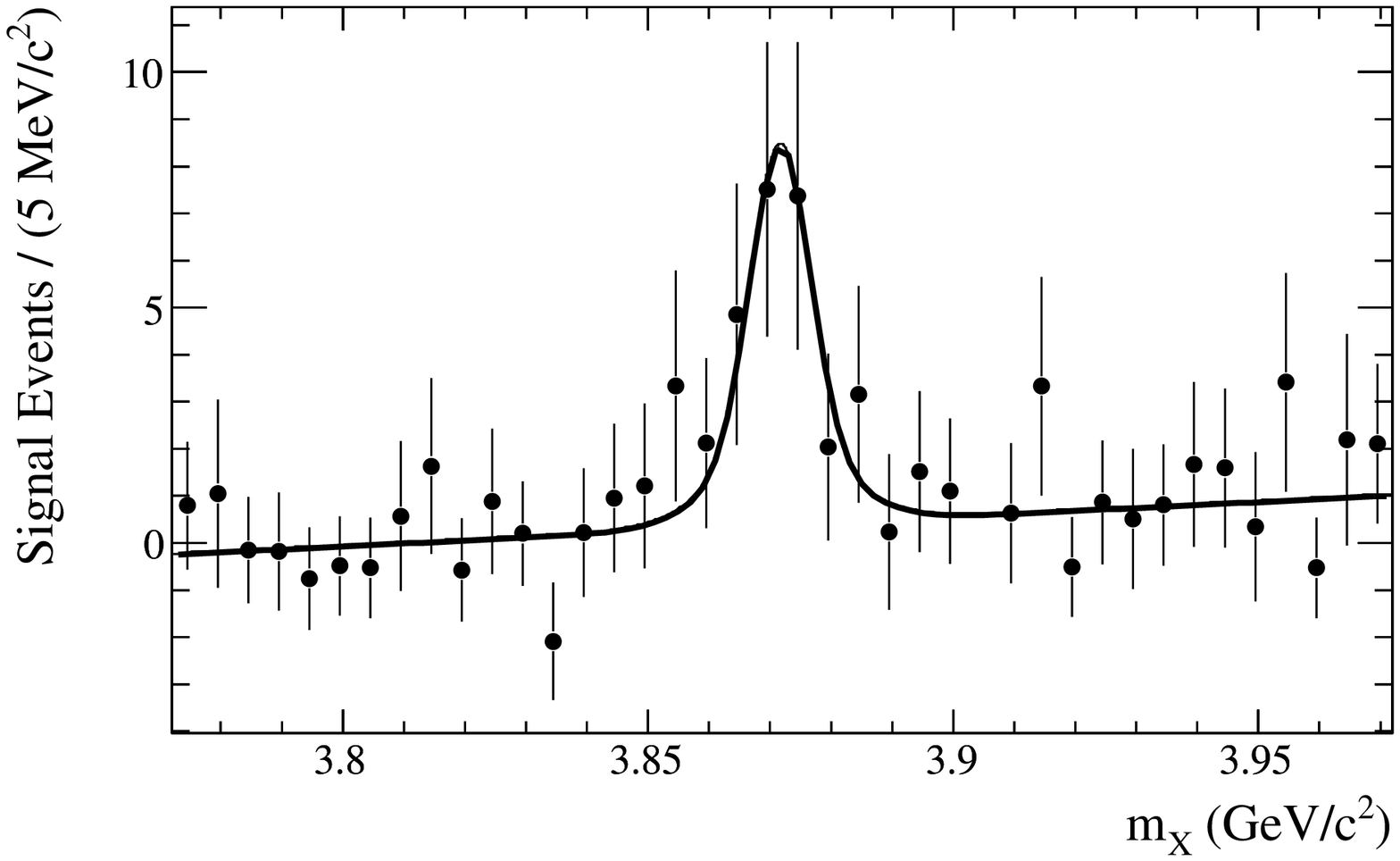} &
\includegraphics[width=0.47\textwidth]{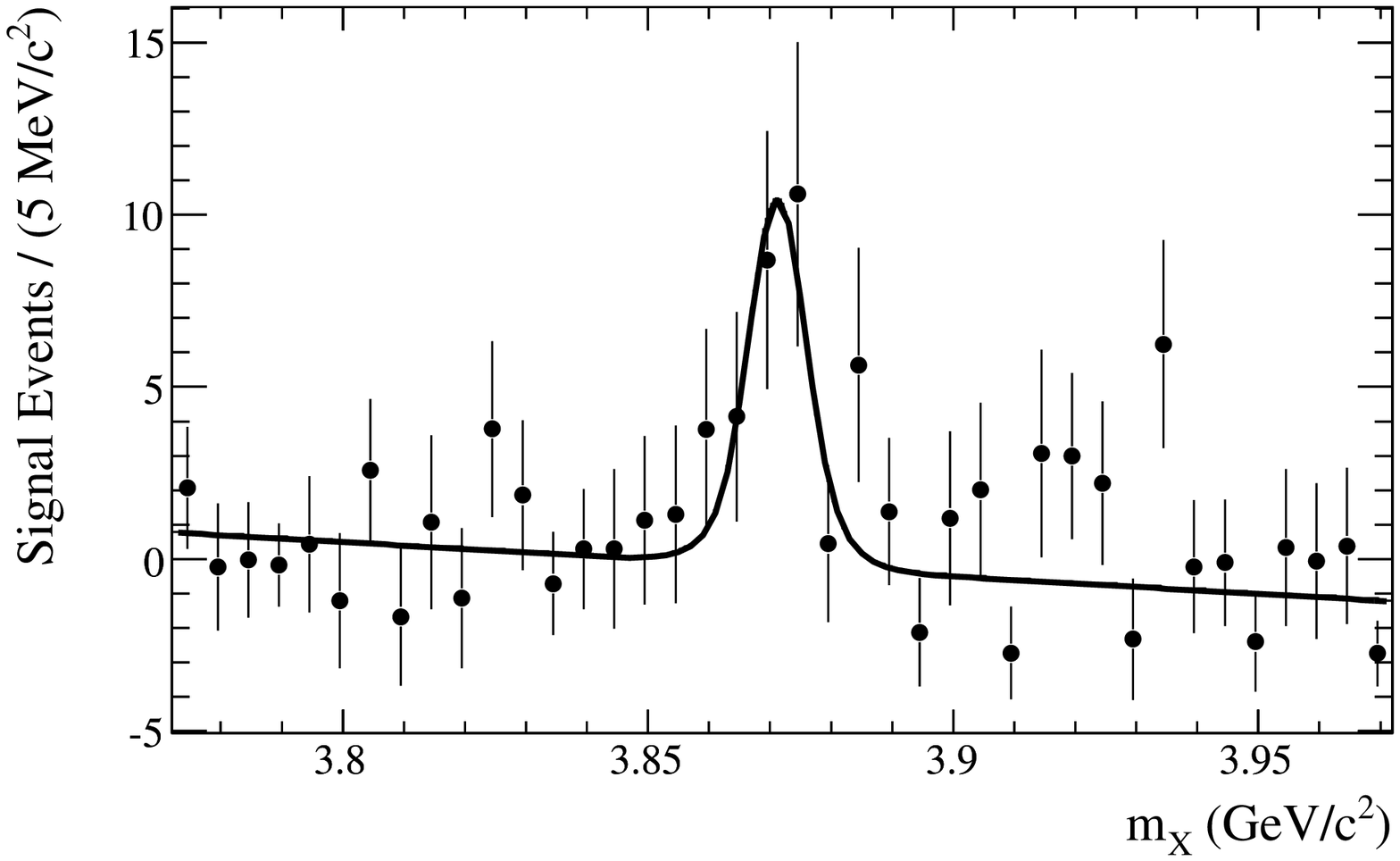}
\end{tabular}
\caption{Babar~\cite{babar:x3872_gam08}: Mass distributions of
  $\jp\gamma$ (left) and of $\pp\gamma$ (right) for $B \to X(3872)K$,
  $X(3872)\to\jp(\pp)\gamma$). The points represent the data and the
  solid curves show the results of a fit to a Breit-Wigner enhancement
  above a linear background.}
\label{fig1}
\end{figure*}

This year BaBar updated the measurement of $X(3872)\to\jp\gamma$ and
reported a new decay mode,
$X(3872)\to\pp\gamma$,~\cite{babar:x3872_gam08}. Figure~\ref{fig1}
shows the $X(3872)$ signals seen in these two modes. The branching
fraction ${\cal{B}}(B^{\pm}\to X(3872)K^{\pm})\times
{\cal{B}}(X(3872)\to\jp\gamma)=(2.8\pm 0.8(\st)\pm 0.2(\sys))\times
10^{-6}$ is in agreement with previous
measurements~\cite{babar:x3872_gam,belle:x3872_gam}, while
${\cal{B}}(B^{\pm}\to X(3872)K^{\pm})\times {\cal{B}}(X(3872)\to
\pp\gamma)=(9.9\pm 2.9(\st)\pm 0.6(\sys))\times 10^{-6}$ is found to
be unexpectedly large. This measurement is inconsistent with a pure
\dndst\ molecule interpretation of $X(3872)$ and favors the model
assuming mixing of a \dndst\ molecule with a conventional charmonium
state~\cite{Swanson:2006}.

As a check of the tetraquark hypothesis, BaBar searched for a charged
partner of the $X(3872)$ in the decay $B\to X(3872)^-K$,
$X(3872)^-\to\jp\pi^-\pi^0$~\cite{babar:x3872_part}. The obtained
upper limits on the production of charged $X(3872)$ partners are
${\cal B}(B^0 \to X(3872)^- K^+) \times {\cal
  B}(X(3782)^-\to\jp\pi^-\pi^0) < 5.4 \times 10^{-6}$ at the $90\%$
C.L.  and ${\cal B }( B^- \to X(3872)^- K^0_S) \times {\cal B
}(X(3872)^- \to \jp \pi^- \pi^0) < 22\times 10^{-6}$ at the $90\%$
C.L. This measurement excludes an isovector hypothesis for the
$X(3872)$.

The diquark-diantiquark model~\cite{x3872:tetra} predicts that the
observed $X(3872)$ is one component of a doublet of states. In this
model, the $X(3872)$ produced in charged $B$ decays would have a mass
that is different from its counterpart in neutral $B$ decays by
$\Delta M = (7 \pm 2)/\cos (2\theta)\mevc$, where $\theta$ is a mixing
angle that is near $\pm 20^{\circ}$. In order to test this hypothesis
both BaBar~\cite{babar:x3872_dmass} and Belle~\cite{belle:x3872_kpi}
performed studies of the $X(3872)$, produced in $B^+ \to X(3872) K^+$
and $B^0 \to X(3872) K^0_S$ decays, where $X(3872) \to \jp \pi^+
\pi^-$. The ratio of branching fractions ${\cal B}(B^0 \to
X(3872)K^0)/{\cal B}(B^+ \to X(3872) K^+)$ is found to be $0.41 \pm
0.24(\st) \pm 0.05(\sys)$ (BaBar) and $0.82 \pm 0.22 (\st) \pm
0.05(\sys)$ (Belle). These values are consistent with unity. The mass
difference between the $X(3872)$ states from charged and neutral $B$
decay modes, $\Delta M \equiv M_{XK^+} - M_{XK^0}$, is found to be
$(2.7\pm 1.6(\st)\pm 0.4(\sys))~\mevc$ (BaBar) and $(0.18 \pm
0.89(\st) \pm 0.26(\sys))\mevc$ (Belle) and is consistent with zero.

\begin{figure}[htbp]
\begin{center}
\begin{tabular}{cc}
\includegraphics[width=0.5\textwidth]{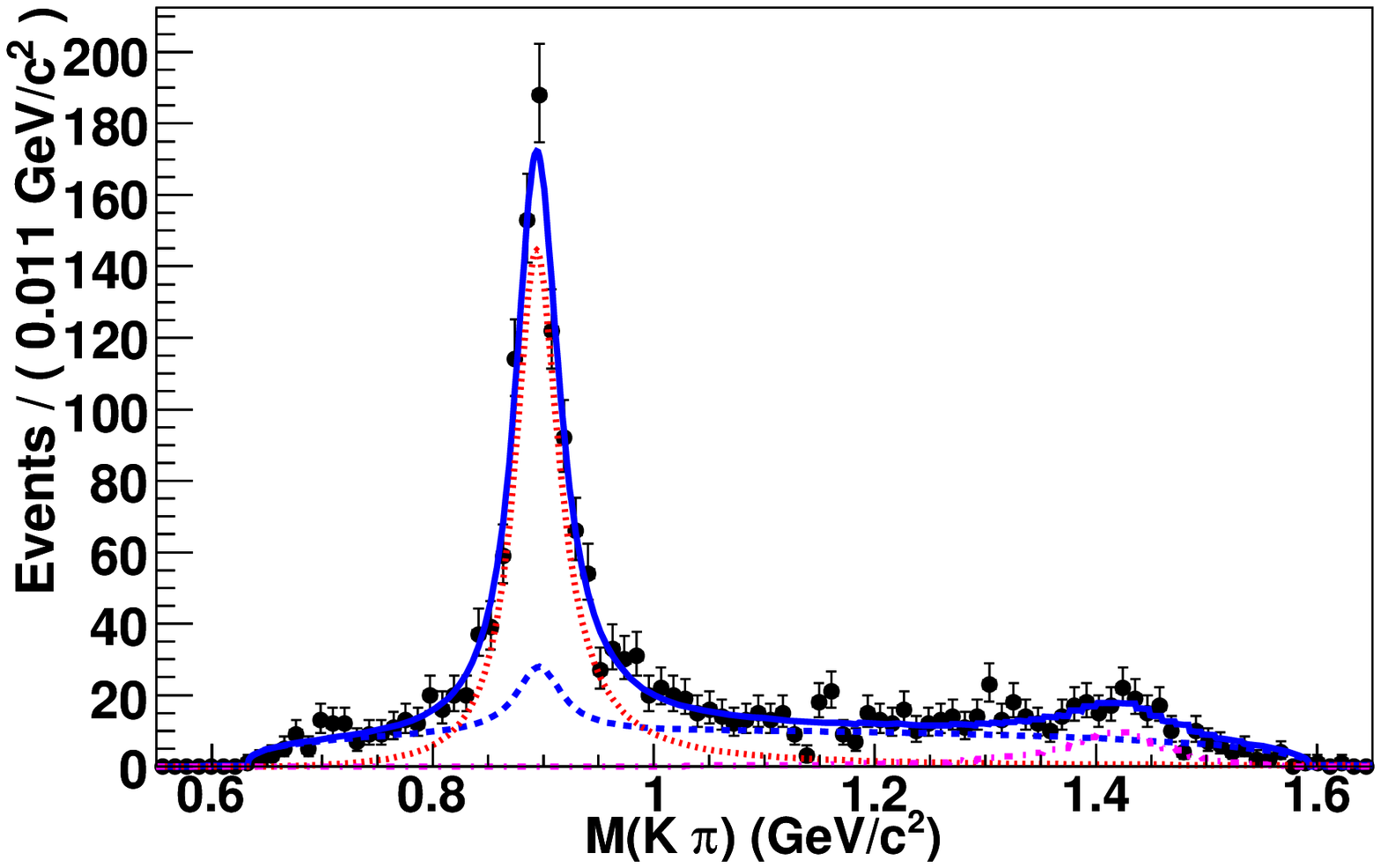} &
\includegraphics[width=0.5\textwidth]{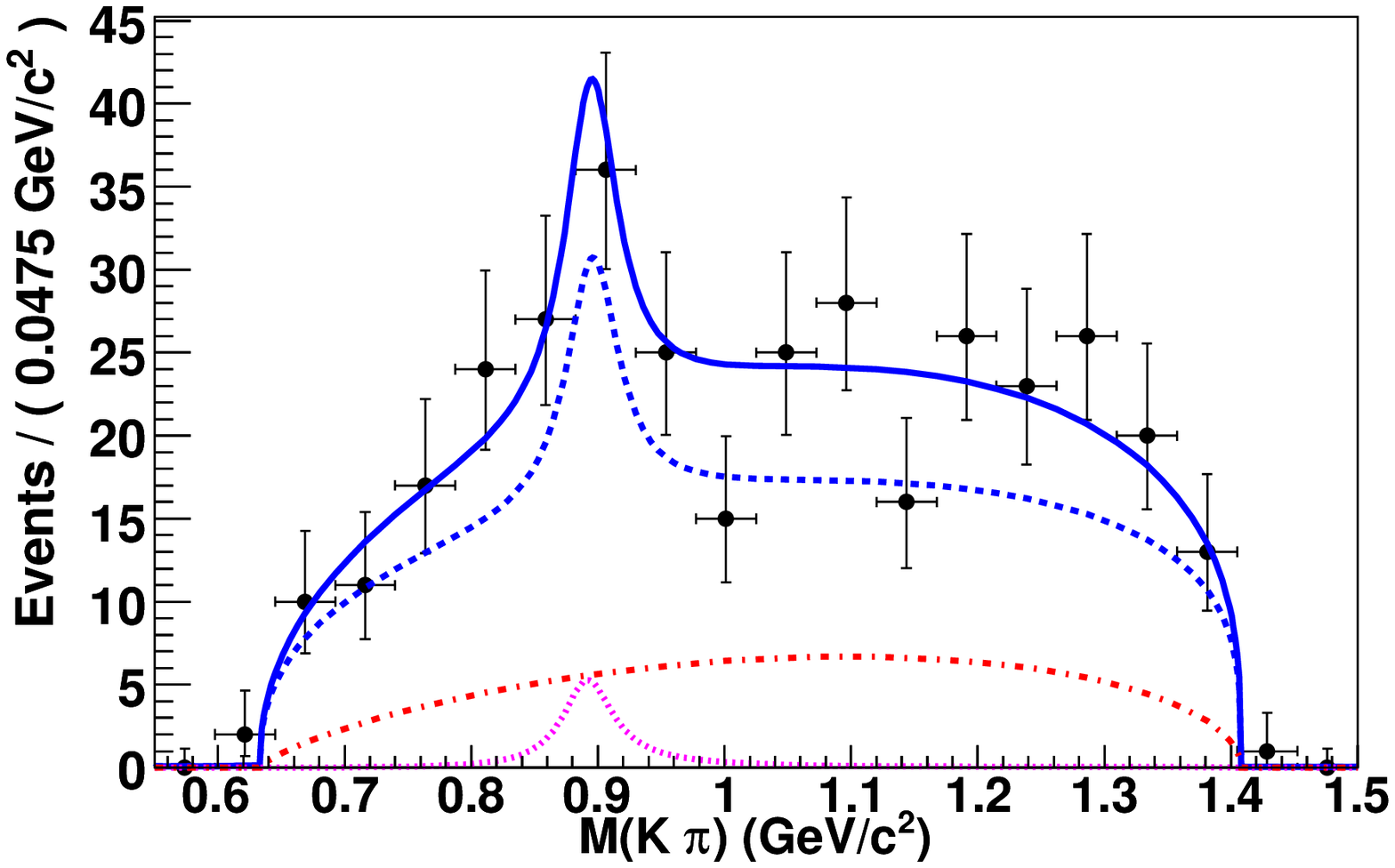}
\end{tabular}
\end{center}
\caption{Belle~\cite{belle:x3872_kpi}: the $K\pi$ mass spectrum for
  the $B\to \pp K \pi$(left) and $B \to X(3872)K \pi$ (right) candidates.
  Left: $B \to\pp K^{*}(892)^0$ is shown by the dotted red curve,
  $B\to\pp K^{*}_2(1430)^0$ by the dash-dot magenta curve, and the
  background by the dashed blue curve. Right: $B \to X(3872) (K^+
  \pi^-)_{\mathrm{nonres}}$ is shown by the dash-dot red curve, $B \to
  X(3872) K^{*}(892)^0$ by the dotted magenta curve, and the
  background by the dashed blue curve.}
\label{fig2}
\end{figure}

In addition, Belle searched for the $X(3872)$ in the decay $B^0 \to
X(3872) K^+ \pi^-$, $X(3872)\to\jp\pipi$~\cite{belle:x3872_kpi} and
measured ${\cal B}(B^0 \to X(3872) (K^+ \pi^-)_{\mathrm{nonres}})
\times {\cal B}(X(3872)\to \jp \pipi) = (8.1 \pm
2.0(\st)^{+1.1}_{-1.4}(\sys))\times 10^{-6}$. Unlike conventional
charmonium the resonant contribution is found to be unexpectedly
small: ${\cal B}(B^0 \to X(3872)K^{*}(892)^0) \times {\cal
  B}(X(3872)\to\jp\pipi)< 3.4\times 10^{-6}$ at the $90\%$ C.L. The
comparison of the $K\pi$ mass spectra for the $B \to \pp K \pi$ and $B
\to X(3872) K \pi$ candidates is shown in Figure~\ref{fig2}.

To test the hypothesis that the $X(3872)$ signal observed in the
\ppjpsi\ decay mode contains two different states,
CDF~\cite{CDF:x3872_mass} performed a careful study of the $X(3872)$
line shape. The resolution function was studied using the \pp\
signal. If two resonances are really merged, their mass splitting was
found to be $\Delta M < 3.2 (3.6)\mevc$ at the $90\%$ ($95\%$) C.L.,
assuming an equal fraction for the two states in the peak. The
measured $X(3872)$ mass value by CDF, $M_X=(3871.61 \pm 0.16(\st)\pm
0.19(\sys))\mevc$, is the most precise mass measurement at the current
time.

\begin{figure*}[t]
\centering
\includegraphics[width=0.9\textwidth]{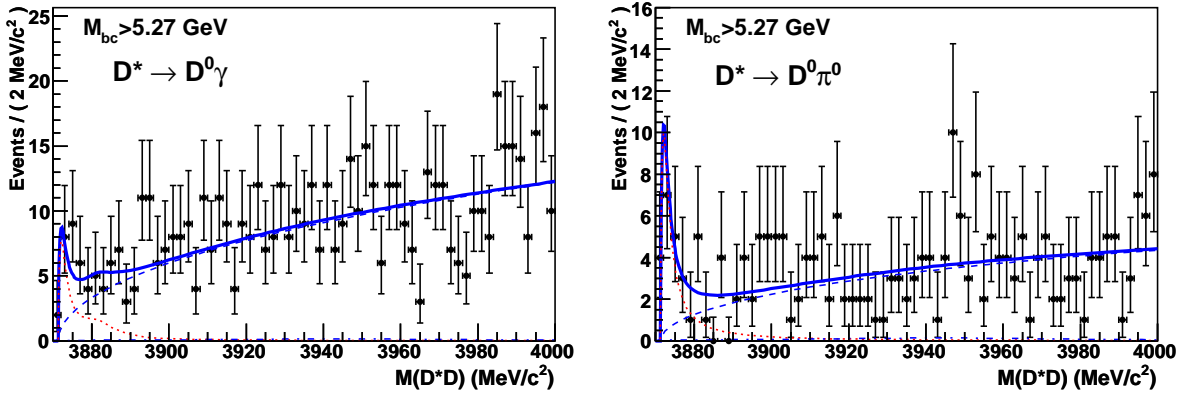}
\caption{Belle~\cite{belle:x3875_3872}:distribution of $M_{\dndst}$
  for $\overline D{}^{*} \to \overline D{}^{0} \gamma$ (left) and
  $\overline D{}^{*} \to \overline D{}^{0} \pi^0$ (right).  The points
  with error bars are data, the dotted curve is the signal, the dashed
  curve is the sum of the background and the $B \to \dndst K$
  component, the dash-dotted curve is the contribution from the
  $Y(3940)$, and the solid curve is the total fitting function.}
\label{fig3}
\end{figure*}

In 2005 Belle showed a $6.4\sigma$ excess of events in the $\dnd
\pi^0$ invariant mass in the channel $B \to \dnd \pi^0 K$, with a mass
of $\sim 3875\mevc$~\cite{belle:x3875} (Table~\ref{tab1}). This year
BaBar also reported an observation of $X(3875)$ decays to
\dndst~\cite{babar:x3875}. The measured mass value, presented in
Table~\ref{tab1}, was found to be in good agreement with the value
measured by Belle and the weighted average is $4.5\sigma$ away from
the mass measured in the $\jp\pipi$ decay mode. This year Belle
presented an updated study of near-threshold enhancement in the
\dndst\ invariant mass spectrum in $B \to \dndst K$ decays
(Figure~\ref{fig3})~\cite{belle:x3875_3872}. The significance of this
enhancement is $8.8\sigma$ and the measured mass and width are
consistent with the current world average values for the $X(3872)$ in
the \ppjpsi\ mode ~\cite{pdg08}. The mass is $2.6\sigma$ lower than
the value obtained by BaBar~\cite{babar:x3875}. The obtained branching
fraction and width are compatible with the values previously published
by Belle in Ref.~\cite{belle:x3875} for non-resonant $\dnd \pi^0$
decays; the mass is $1\sigma$ lower, while only $30\%$ of the data
sample is in common. An alternative fitting method, using the Flatt\'e
distribution, gives results similar to those obtained with a
traditional Breit-Wigner function.
\begin{table}[htb]
\caption{Measured parameters of the $X(3872)$ state}
\label{tab1}
\begin{center}
\footnotesize
\begin{tabular}{c|c|c|c|c|c|c}
\hline \hline
State & $M,\mevc$ & $\gtot,\mev$ & $J^{PC}$ & Decay Modes & Production &
Collaboration\\\hline
$X(3872)$&$3872.8\pm0.8$&$3.0^{+2.1}_{-1.7}$&$1^{++}$&$\pipi\jp$& $B\to
KX(3872)$
& Belle, BaBar\\
& PDG08 &  PDG08 &$1^{++}$&$\pipi\jp$&$B\to X(3872)X$, $p\overline{p}$ &
 CDF, D0\\\hline
$X(3875)$&$3875.2\pm
0.7^{+0.9}_{-1.8}$&$1.22\pm0.31^{+0.23}_{-0.30}$&$?^{??}$&
$\dnd \pi^0$&$B\to K X(3875)$& Belle~\cite{belle:x3875} \\
$X(3875)$&$3875.1^{+0.7}_{-0.5}\pm 0.5$&$3.0 {}^{+1.9}_{-1.4} \pm
0.9$&$?^{??}$ &
\dndst & $B\to K X(3875)$& BaBar~\cite{babar:x3875}\\
$X(3872)$& $3872.6^{+0.5}_{-0.4}\pm 0.4
$&$3.9^{+2.5}_{-1.3}{}^{+0.5}_{-0.3}$&
$?^{??}$&$\dnd\pi^0(\gamma)$&$B\to K X(3872)$&
Belle~\cite{belle:x3875_3872} \\
\hline\hline
\end{tabular}
\end{center}
\end{table}

This year we celebrate the fifth anniversary of the $X(3872)$!  In
spite of a large accumulation of experimental data and numerous
theoretical approaches, the nature of this mysterious state remains to
be established.

%%%%%%%%%%%%%%%%%%%%%%%%%%%%%%%%%%%%%%%%%%%%%%%%%%%%%%%%%%%%%%%%%%%%%%%%%%%

\section{ISR family}

The whole family of unexpected charmoniumlike states with masses above
open charm threshold were discovered in the $\ee \to
\ppjpsi(\pp)\gisr$ processes. (Measured masses and widths of discussed
$Y$ states are presented in Table~\ref{tab2}.) The production with
initial-state radiation (ISR) fixes the quantum numbers of these
states to be $J^{PC}=1^{--}$. The absence of open charm decay channels
for these states is inconsistent with their interpretation as
conventional charmomium. Partial widths of decay channels to
charmonium plus light hadrons are found to be much larger than those
usual for conventional charmonium states.

\begin{figure*}[t]
\centering
\begin{tabular}{cc}
\includegraphics[width=0.47\textwidth]{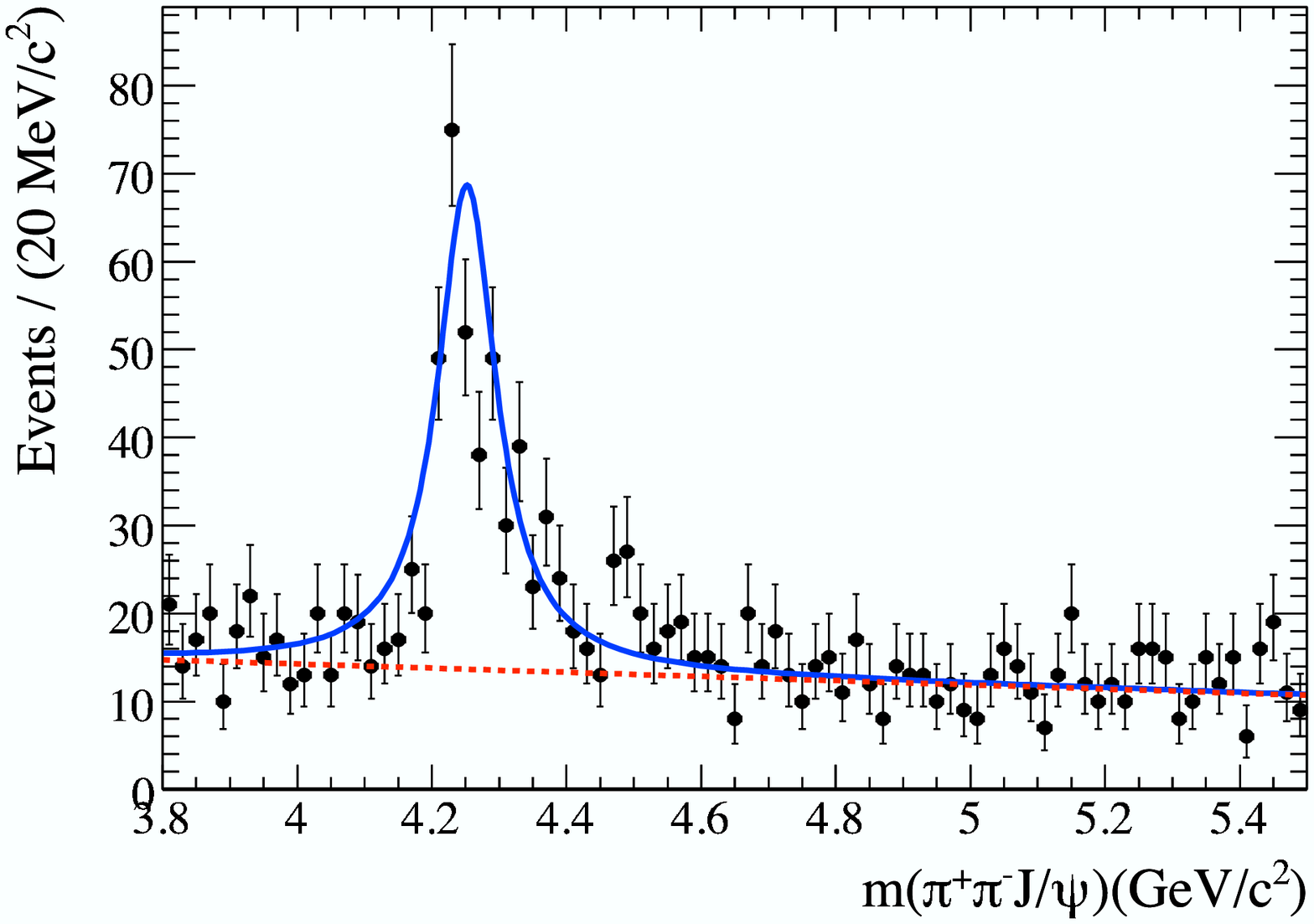} &
\includegraphics[width=0.5\textwidth]{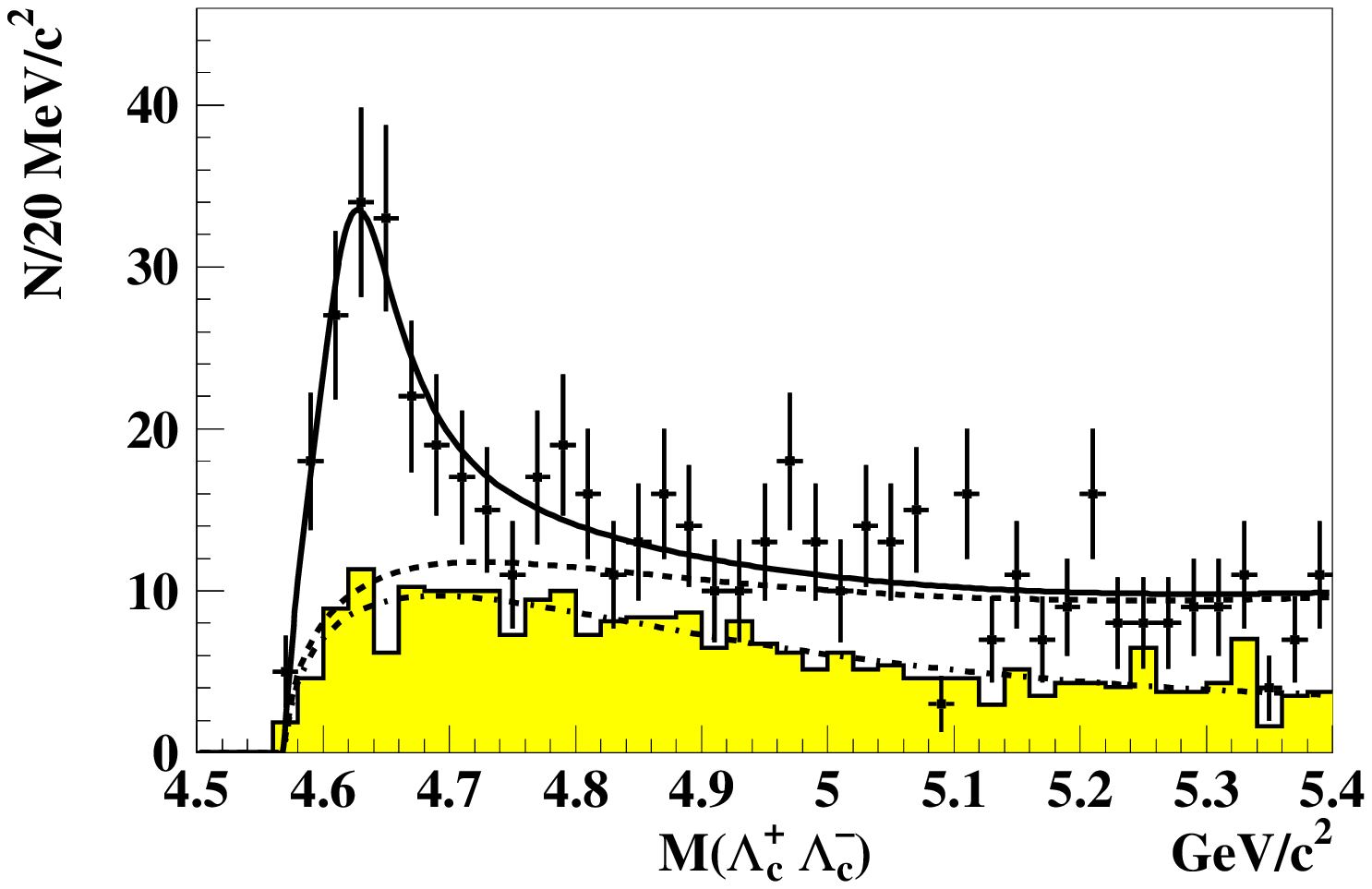}
\end{tabular}
\caption{(Left) BaBar~\cite{babar:y4260_08}: the ISR-produced
  \ppjpsi\ invariant mass distribution. The dots represent the data,
  the solid curve shows the fit result, while the dashed curve
  represents the background contribution. (Right)
  Belle~\cite{belle:x4630}: the \mll\ spectrum for the signal
  region. The solid curve represents the result of the fit. The
  threshold function is shown by the dashed curve. The combinatorial
  background parameterizations is shown by the dashed-dotted curve.}
\label{fig4}
\end{figure*}

The first state, called the $Y(4260)$, was discovered by BaBar as an
accumulation of events near $4.26\gevc$ in the invariant mass spectrum
of \ppjpsi~\cite{babar:y4260}. The new resonance was immediately
confirmed by CLEO both in a similar ISR study using data taken in the
$\Upsilon(1S-4S)$ region~\cite{cleo:y4260_isr} and in the energy scan
at $\sqs =(3.97-4.26)\gevc$~\cite{cleo:y4260_scan}. In addition to the
$Y(4260)$ state observed in the \ppjpsi\ invariant mass distribution,
Belle has found another wide cluster of events around $4.0\gevc$,
called the $Y(4008)$~\cite{belle:y4260}. It has been shown that a fit
using two interfering Breit-Wigner amplitudes described the data
better than a fit assuming one resonance, especially for the
lower-mass side of the $4.26\gevc$ enhancement. This year BaBar
presented an update of the $Y(4260)$ resonance study
(Figure~\ref{fig4}, left) and did not confirm the broad structure
around $4.0\gevc$~\cite{babar:y4260_08}.

Another structure, called the $Y(4325)$, was observed by BaBar in the
$\ee \to \pppsip$ cross-section near $4.32\gevc$~\cite{babar:y4360}.
Belle performed a similar study and claimed the existence of two
resonant structures, one, in agreement with the BaBar study, is
observed near $4.36\gevc$ and another, called $Y(4660)$, near
$4.66\gevc$.  Belle parameterized the \pppsip\ mass spectrum with a
coherent sum of two Breit-Wigner amplitudes~\cite{belle:y4360}. No
sign was found either of $Y(4260)$ ($Y(4008)$) decay to \pppsip, or of
$Y(4325)$ ($Y(4660)$) decay to \ppjpsi.

The observation of the $Y(4260)$ motivated numerous measurements of
exclusive \ee\ cross sections for charmed hadron pairs near
threshold. Belle presented the first results of the exclusive
\ee\ cross sections to \dd\ ($D=D^0$ or $D^+$), \dpdstm,
\dstpdstm\ and \dndmpp\ final states using
ISR~\cite{belle:dd,belle:dst,belle:4415}. BaBar has measured cross
sections for \eedd\ using ISR~\cite{babar:dd}. CLEO-c performed a scan
over the energy range from 3.97 to 4.26\gev\ and measured exclusive
cross sections for \dd, \ddst\ and \dstdst\ final states at twelve
points with high accuracy~\cite{cleo:cs}. Surprisingly, no evidence
for open charm production associated with any of the $Y$ states (which
is expected for a conventional charmonium with such large masses and
widths) has been observed, in fact the $Y(4260)$ peak position appears
to be close to a local minima of both the total hadronic cross
section~\cite{bes:cs} and of the exclusive cross section for
$\ee\to\dstpdstm$~\cite{belle:dst}.

The absence of available $J^{PC}=1^{--}$ charmonium levels for the $Y$
states is another problem for their
interpretation. $\psi(4040)=3^3S_1$, $\psi(4160)=2^3D_1$ and
$\psi(4415)=4^3S_1$ states are well established, while masses of
$3^3D_1(4560)$, $5^3S_1(4760)$, $4^3D_1(4810)$ states predicted by the
quark model~\cite{godfrey_isgur} are higher or lower than the observed
$Y$ masses. To resolve this problem some models calculate
$3^3D_1(4350)$, $5^3S_1(4660)$ with shifted
masses~\cite{y:charm}. Coupled-channel effects and re-scattering of
charm mesons is another possible way to explain the observed
peaks~\cite{voloshin:rescattering}. The most popular exotic option for
the $Y$ states is that they are the hybrids expected by LQCD at
$4.2-5.0\gevc$~\cite{y4260:hybryds}. Other suggestions are
hadro-charmonium~\cite{hadroch}; multiquark states including
$[cq][\overline{cq}]$ tetraquark~\cite{y4260:tetra}, $D \overline
D{}_1$ or \dndst\ molecules~\cite{y4260:molecule}, $f_0(980)\pp$
molecule for the $Y(4660)$~\cite{y4660:molecule} .

Recently Belle has reported a significant near-threshold enhancement
in the \eell\ exclusive cross section, called the $X(4630)$
(Figure~\ref{fig4}, right)~\cite{belle:x4630}. The nature of the
$X(4630)$ remains unclear. Assuming the $X(4630)$ to be a resonance,
Belle obtained the mass and width values, presented in
Table~\ref{tab2}, which are consistent within errors with the mass and
width of the $Y(4660)$ state. On the other hand, in many processes,
including three-body $B$ meson baryonic decays, mass peaks are
observed near baryon antibaryon mass
thresholds~\cite{dibaryon}. Finally, the interpretation of the
$X(4630)$ as a conventional $5^3S_1$ charmonium
state~\cite{x4630:charm,y:charm} or as a threshold effect, caused by
the presence of the $\psi(3D)$ just below the \lala\ threshold cannot
be excluded~\cite{x4630:psi3s}.
\begin{table}[htb]
\caption{Measured parameters of the $Y$ states}
\label{tab2}
\begin{center}
\footnotesize
\begin{tabular}{c|c|c|c|c|c|c}
\hline \hline
State & $M,\mevc$ & $\gtot,\mev$ & $J^{PC}$ & Decay Modes &
Production & Collaboration\\\hline
$Y(4008)$&$4008\pm 40^{+114}_{-28}$ & $226\pm 44\pm 87$   &$1^{--}$
&$\pipi\jp$
&$\ee$(ISR)&Belle 07~\cite{belle:y4360}\\\hline
$Y(4260)$&$4259\pm 8^{+2}_{-6}$     & $88\pm 23 ^{+6}_{-4}$&$1^{--}$
&$\pipi\jp$
&$\ee$(ISR)&BaBar 05~\cite{babar:y4260}\\
$Y(4260)$&$4252\pm 6^{+2}_{-3}$&$105\pm 18 ^{+4}_{-6}$    &$1^{--}$
&$\pipi\jp$
&$\ee$(ISR)&BaBar 08~\cite{babar:y4260_08}\\
$Y(4260)$&$4247\pm 12^{+17}_{-32}$&$108\pm 19\pm 10$      &$1^{--}$
&$\pipi\jp$ &
$\ee$(ISR)&Belle 07~\cite{belle:y4260}\\\hline
$Y(4325)$&$4324\pm 24$& $172\pm33$                       &$1^{--}$
&$\pipi\pp$ &
$\ee$(ISR)&BaBar 06~\cite{babar:y4360}\\
$Y(4325)$&$4361\pm 9\pm 9$&$74\pm 15\pm 10$              &$1^{--}$
&$\pipi\pp$ &
$\ee$(ISR)&Belle 07~\cite{belle:y4360}\\\hline
$Y(4660)$&$4664\pm 11\pm 54$&$48\pm 15\pm 3$             &$1^{--}$
&$\pipi\pp$ &
$\ee$(ISR)&Belle 07~\cite{belle:y4360}\\\hline
$X(4630)$&
$4634^{+8}_{-7}{}^{+5}_{-8}$&$92^{+40}_{-24}{}^{+10}_{-21}$&$1^{--}$&\lala
&
$\ee$(ISR)&Belle 08~\cite{belle:x4630}\\\hline\hline
\end{tabular}
\end{center}
\end{table}

%%%%%%%%%%%%%%%%%%%%%%%%%%%%%%%%%%%%%%%%%%%%%%%%%%%%%%%%%%%%%%%%%%%%%%%%%%%%%%%%%

\section{The $XYZ(3940)$ family}

Curiously, three charmoniumlike states were observed with similar
masses near $3.94\gevc$, but in quite different processes
(Table~\ref{tab3}). The $Z(3930)$ state was found by Belle in
two-photon collisions $\gamma \gamma \to \dd$ with a mass $\sim
3.930\gevc$~\cite{belle:z3930}. The production rate and the angular
distribution in the $\gamma \gamma$ center-of-mass frame favor the
interpretation of $Z(3930)$ as the $\chi_{c2}(2P)$ charmonium state.

\begin{figure}[htb]
\begin{center}
\begin{tabular}{cc}
\includegraphics[width=0.47\textwidth]{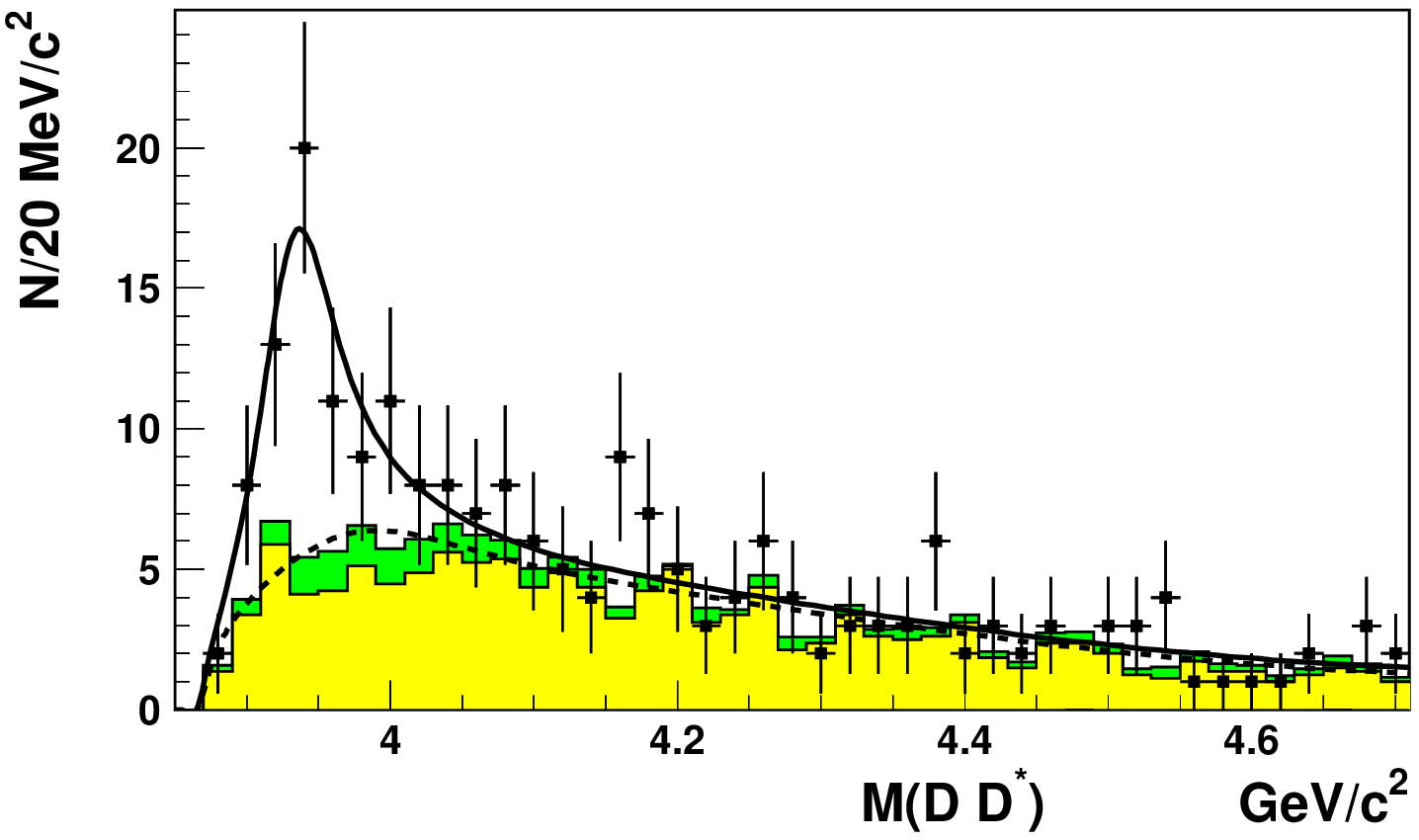} &
\includegraphics[width=0.47\textwidth]{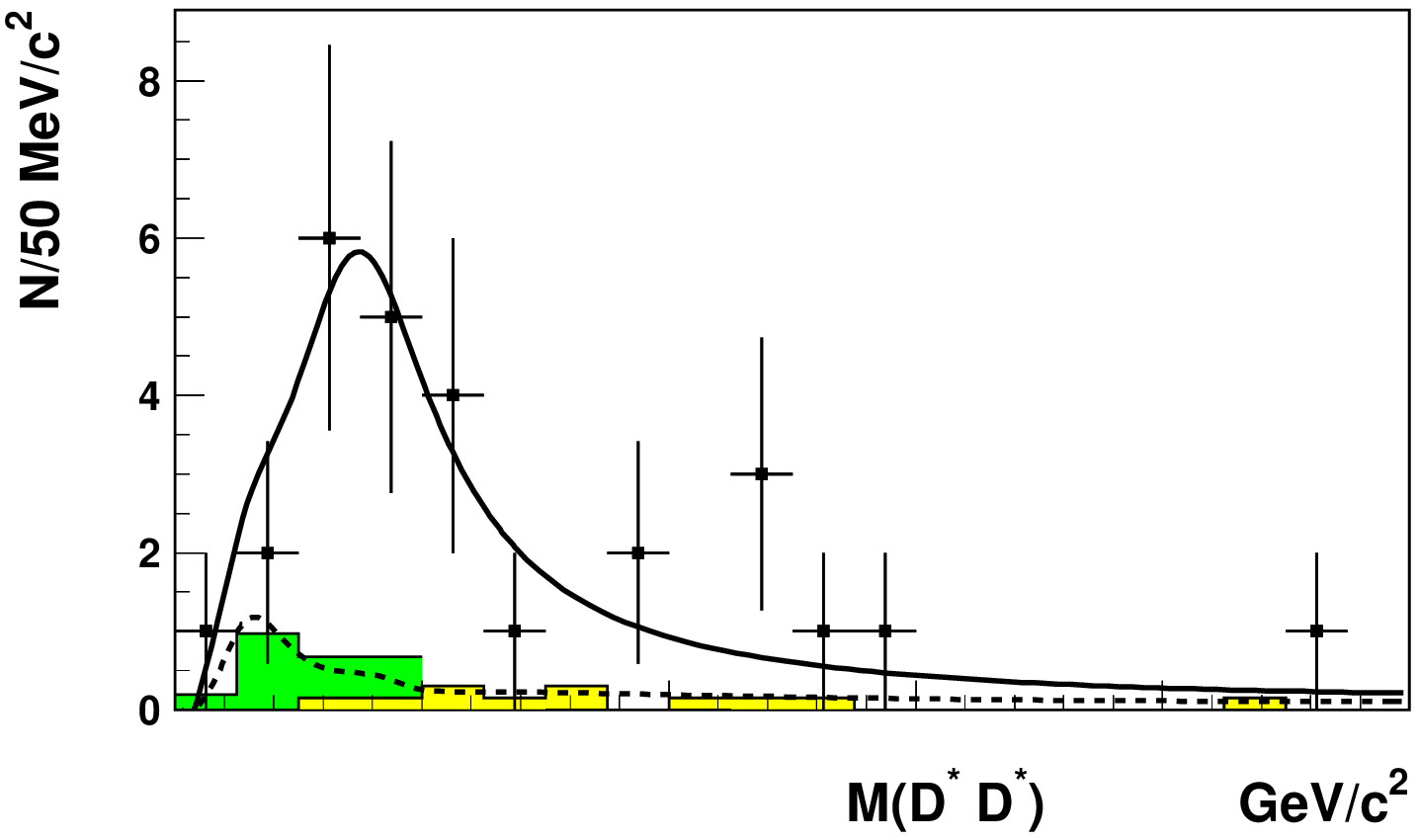}
\end{tabular}
\caption{Belle~\cite{belle:x4160}: The $M(\ddst)$ spectra for the
  $\ee\to \jp\ddst$ process (left). The $M(\dstdst)$ spectra for the
  $\ee\to \jp\dstdst$ process (right). Points with error bars
  correspond to the signal windows while histograms show the scaled
  sideband distributions.}
 \label{fig5}
\end{center}
\end{figure}

Another charmoniumlike state, the $X(3940)$, has been observed by
Belle in the double charmonium production\footnote{The double
  charmonium production in \ee\ annihilation was first observed by
  Belle in 2002~\cite{belle:2cc_first}. It was also found that scalar
  and pseudoscalar charmonia are produced copiously recoiling against
  \jp\ or \pp.} in the process $\ee \to \jp \ddst$ in the mass
spectrum recoiling against the \jp~\cite{belle:x3940}. This year Belle
confirmed the observation of $X(3940)\to \ddst $ with a significance
of $5.7\sigma$ (Figure~\ref{fig5})~\cite{belle:x4160}. In addition
Belle found a new charmoniumlike state, $X(4160)$, in the processes
$\ee \to \jp X(4160)$ decaying into \dstdst\ with a significance of
$5.1\sigma$. Parameters of the $X(3940)$ and $X(4160)$ are presented
in Table~\ref{tab3}. Both the $X(3940)$ and the $X(4160)$ decay to
open charm final states and therefore could be attributed to $3^1S_0$
and $4^1S_0$ conventional charmonium states. However, the problem with
this assignment is that potential models predict masses for these
levels to be significantly higher than those measured for the
$X(3940)$ and $X(4160)$.

Finally, the $Y(3940)$ state was observed by Belle as a near-threshold
enhancement in the $\omega\jp$ invariant mass distribution for
exclusive $B\to K \omega \jp$ decays with a statistical significance
of $8.1\sigma$~\cite{belle:y3940}. Recently BaBar found an $\omega\jp$
mass enhancement at $\sim 3.915\gevc$ in the decays $B^{0,+} \to
K^{0,+} \omega \jp$~\cite{babar:y3940} and confirmed the Belle result,
but obtained a lower mass and smaller width and reduced the
uncertainty on each by a factor of $\sim 3$. The $Y(3940)$ mass is two
standard deviations lower than the $Z(3930)$ mass, and three standard
deviations lower than for the $X(3940)$; the width agrees with the
$Z(3930)$ and $X(3940)$ values. The ratio of $B^0$ and $B^+$ decay to
$Y(3940)K$, $R_Y = 0.27^{+0.28}_{-0.23}(\st)^{+0.04}_{-0.01}(\sys)$,
is found to be $\sim 3$ standard deviations below the isospin
expectation, but agrees with that for the
X(3872)~\cite{babar:x3872_dmass}.

In the Belle study of the $\ee \to \jp X(3940)$
process~\cite{belle:x3940}, no signal for the decay $X(3940)\to
\omega \jp$ was found.  On the other hand the decay $Y(3940)\to
\dndst$ was not observed in the $B\to
Y(3940)K$~\cite{belle:x3875_3872}. The obtained upper limits: ${\cal
  B}(Y(3940)\to \omega \jp)/{\cal B}(Y(3940) \to \dndst) > 0.71$ at
the $90\%$ C.L.  and ${\cal B}(X(3940) \to \omega \jp){\cal B}(X(3940)
\to \dndst)< 0.58$ at the 90\% C.L., allows to claim that the
$X(3940)$ and the $Y(3940)$ are different
states~\cite{belle:x3875_3872}.

\begin{table}[htb]
\caption{Measured parameters of the $XYZ(3940)$ states}
\label{tab3}
\begin{center}
\footnotesize
\begin{tabular}{c|c|c|c|c|c|c}
\hline \hline
State & $M,\mevc$ & $\gtot,\mev$ & $J^{PC}$ & Decay Modes &
Production & Collaboration \\\hline
%%%% cccc %%%%%
$X(3940)$&$3942^{+7}_{-6}\pm 6$&$37^{+26}_{-18}\pm8$       &$?^{?+}$&\ddst
&$\ee\to\jp X(3940)$& Belle 07~\cite{belle:x4160}\\
$X(4160)$&$4156^{+25}_{-20}\pm15$&$139^{+111}_{-61}\pm 21$
&$?^{?+}$&\dstdst
&$\ee\to\jp X(4160)$& Belle 07~\cite{belle:x4160}\\\hline
%%%%% Y(3940) %%%%%
$Y(3940)$&$3943\pm 11\pm 13$&$87\pm 22\pm26$&$?^{?+}$&$\omega\jp$&
$B\to K Y(3940)$&Belle 05~\cite{belle:y3940}\\
$Y(3940)$&$3914.6^{+3.8}_{-3.4}\pm 2.0$&$34^{+12}_{-8}\pm 5$&$?^{?+}$&
$\omega\jp$&$B\to K Y(3940)$&BaBar 08~\cite{babar:y3940}\\\hline
%%%%% Z(3930) %%%%%
$Z(3930)$& $3929\pm 5\pm 2$&$29\pm 10\pm 2$&$2^{++}$& \dd  &
$\gamma\gamma\to Z(3940)$& Belle 05~\cite{belle:z3930}\\
\hline\hline
\end{tabular}
\end{center}
\end{table}

%%%%%%%%%%%%%%%%%%%%%%%%%%%%%%%%%%%%%%%%%%%%%%%%%%%%%%%%%%%%%%%%%%%%

\section{The first candidates for charmonumlike states with nonzero electric
charge}

Last year Belle reported an observation of the first candidate
charmoniumlike state with nonzero electric
charge~\cite{belle:z4430}. Such a state, if it exists, could only be a
multiquark state and not conventional charmonium or a hybrid. A
distinct peak, called the $Z(4430)^+$, was found in the $\pi^{\pm}\pp$
invariant mass distribution near $4.43\gevc$ in $B\to K \pi^{\pm}\pp$
decays with a statistical significance of $6.5\sigma$. A fit using a
Breit-Wigner resonance shape yields a peak mass and width of $(4433
\pm 4 (\st) \pm 2(\sys))\mevc$ and $\Gamma = (45^{+18}_{-13}
(\st)^{+30}_{-13} )\mev$. The product branching fraction is determined
to be ${\cal B}(B\to K Z(4430)) \times {\cal B}(Z(4430)\to \pi^+
\pp)=(4.1\pm 1.0(\st)\pm 1.4(\sys))\times 10^{-5}$.

This year BaBar presented a search for $Z(4430)^- \to \jp \pi^-$ and
$Z(4430)^- \to \pp \pi^-$ in $B \to K \pi^- \jp (\pp)$ decays, where
$K=K^0_S$ or $K^+$~\cite{babar:z4430}. BaBar performed a detailed
study of $K\pi^-$ reflections into the $\jp\pi^-$ and $\pp\pi^-$
masses (in S-, P-, D-waves) to describe background for both \jp\ and
\pp\ modes. From the fits to $\jp\pi^-$ mass distribution, in which
background shape was fixed and S-wave Breit-Wigner is used as signal
function, no evidence for any enhancement for \jp\ samples was
obtained. From a similar fit for \pp\ data, small signals with
significance less than $3\sigma$ were obtained. BaBar claimed no
significant evidence for existence of the $Z(4430)^-$.

Recently, Belle reported the first observation of two resonance-like
structures in the $\pi^+ \chi_{c1}$ invariant mass distribution near
$4.1\gevc$ in exclusive $\overline B{}^0\to K^-\pi^+\chi_{c1}$
decays~\cite{belle:z12}. From a Dalitz plot analysis in which the
$\pi^+ \chi_{c1}$ mass structures are represented by Breit-Wigner
resonance amplitudes, Belle determined masses and widths of these new
structures, presented in Table~\ref{tab4}, and product branching
fractions of ${\cal B}(\overline B{}^0 \to K^- Z^+_{1,2})\times{\cal
  B}( Z^+_{1,2}\to\pi^+ \chi_{c1}) =
(3.0^{+1.5}_{-0.8}(\st)^{+3.7}_{-1.6}(\sys))\times 10^{-5}$ and
$(4.0^{+2.3}_{-0.9}(\st)^{+19.7}_{-0.5}(\sys))\times 10^{-5}$,
respectively. The significance of each of the $\pi^+ \chi_{c1}$
structures exceeds $5\sigma$, including the effects of systematics
from various fit models.

To confirm the $Z^{\pm}(4430)$ state more statistics and advanced
Dalitz analysis are needed, while for a confident claim of the
$Z^{\pm}_{1,2}$ existence a confirmation from BaBar is critical.

\begin{table}[htb]
\caption{Measured parameters of the $Z^{\pm}$ states}
\label{tab4}
\begin{center}
\footnotesize
\begin{tabular}{c|c|c|c|c|c|c}
\hline \hline
State & $M,\mevc$ & $\gtot,\mev$ & $J^{PC}$ &
Decay Modes & Production & Collaboration\\\hline
$Z^{\pm}(4430)$&$4433\pm 4\pm
2$&$45^{+18}_{-13}{}^{+30}_{-13}$&$?^{??}$&$\pi^{\pm}\pp$&
$B\to KZ^{\pm}(4430)$&Belle 07~\cite{belle:z4430}\\
$Z^{\pm}_1$&$4051\pm14{}^{+20}_{-41}$&$82^{+21}_{-17}{}^{+47}_{-22}$&$?^{??}$&
$\pi^{\pm}\chi_{c1}$&$B\to KZ^{\pm}_1$&Belle 08~\cite{belle:z12}\\
$Z^{\pm}_2$&$4248^{+44}_{-29}{}^{+180}_{-35}$&$177^{+54}_{-39}{}^{+316}_{-61}$&$?^{??}$&
$\pi^{\pm}\chi_{c1}$&$B\to KZ^{\pm}_2$& Belle 08~\cite{belle:z12}\\
\hline\hline
\end{tabular}
\end{center}
\end{table}

\section{Conclusion}

The discovery of numerous charmoniumlike states discussed in this
review became possible due to the excellent performance of both the
KEKB and PEPII $B$-factories. A very high luminosity of
\ee\ colliders, constructed to search for $CP$-violation in $B$
mesons, allowed them to be used as {\em charmonium} factories. Using
all possible charmonium production mechanisms at center-of-mass energy
near $10.548\gev$, the Belle and BaBar collaborations have made
significant contributions to charmonium spectroscopy. Surprisingly,
the major fraction of the observed charmoniumlike states with masses
above open charm threshold cannot be explained as conventional
charmonium. Although the number of exotic theoretical interpretations
of these states is growing, they can still not explain all of the
existing observations. The field remains puzzling and intriguing. More
efforts are needed both to improve the theoretical understanding and
to perform more precise measurements of exotic states at Super
$B$-factories.

\begin{acknowledgments}

The author would like to thank Belle, BaBar, BES, CLEO, CDF and D0
collaborations for interesting results presented in this review. I am
very thankful to S.A.\,Prell, A.G.\,Mokhtar, P.\,Grenier, D.\,Asner
and M.\,Paulini for helpful communications. It is my great pleasure to
thank the Belle colleagues S.-K.\,Choi, S.\,Olsen, P.\,Pakhlov,
S.\,Uehara, R.\,Mizuk, R.\,Chistov, K.\,Trabelsi, N.\,Zwahlen,
T.\,Aushev, C.Z.\,Yuan, X.L.\,Wang, C.P.\,Shen, P.\,Wang, S.\,McOnie
and the whole Belle collaboration for their recent discoveries and
brilliant studies of charmoniumlike states. I am very grateful to
T.\,Browder, Y.\,Sakai and P.\,Chang, P.\,Pakhlov, S.\,Olsen,
S.\,Eidelman, B.\,Golob and M.\,Kreps for suggestions and comments.

I would like to thank S.\,Smith, J.\,Kroll and N.\,Lockyer for the
invitation to be a plenary speaker at ICHEP08. The author is very
grateful RFBR for financial support (grants RFBR-08-02-08473,
RFBR-1329.2008.2).

\end{acknowledgments}

%%%%%%%%%%%%%%%%%%%%%%%%%%%%%%%%%%%%%%%%%%%%%%%%%%%%%%%%%%%%%%%%%%%%

\end{document}